\documentclass[aps,prb,amsmath,amssymb,reprint,superscriptaddress,preprintnumbers,showpacs,intlimits,longbibliography]{revtex4-1}
\usepackage{bm,latexsym,mathrsfs,enumerate,color}
\usepackage[mathcal]{euscript}
\usepackage[breaklinks=true,unicode=true,urlcolor = blue,colorlinks = true,citecolor = blue,linkcolor = blue]{hyperref}
\usepackage{graphicx}
\usepackage[rightcaption]{sidecap}
\renewcommand{\vec}[1]{\bm{#1}}
\usepackage[normalem]{ulem}

\begin{document}
%
%
\title{Multiplet of skyrmion states on a curvilinear defect: skyrmion lattices as a ground state}

\author{Volodymyr P. Kravchuk}
\email{vkravchuk@bitp.kiev.ua}
\affiliation{Bogolyubov Institute for Theoretical Physics of National Academy of Sciences of Ukraine, 03680 Kyiv, Ukraine}
\affiliation{Leibniz-Institut f{\"u}r Festk{\"o}rper- und Werkstoffforschung, IFW Dresden, D-01171 Dresden, Germany}

\author{Denis D. Sheka}
\affiliation{Taras Shevchenko National University of Kyiv, 01601 Kyiv, Ukraine}

\author{Oleksii M. Volkov}
\affiliation{Helmholtz-Zentrum Dresden-Rossendorf e.V., Institute of Ion Beam Physics and Materials Research, 01328 Dresden, Germany}

\author{Ulrich~K.~R{\"o}{\ss}ler}
\affiliation{Leibniz-Institut f{\"u}r Festk{\"o}rper- und Werkstoffforschung, IFW Dresden, D-01171 Dresden, Germany}

\author{Jeroen~van~den~Brink}
\affiliation{Leibniz-Institut f{\"u}r Festk{\"o}rper- und Werkstoffforschung, IFW Dresden, D-01171 Dresden, Germany}
\affiliation{Institute for Theoretical Physics, TU Dresden, 01069 Dresden, Germany}

\author{Denys Makarov}
\affiliation{Helmholtz-Zentrum Dresden-Rossendorf e.V., Institute of Ion Beam Physics and Materials Research, 01328 Dresden, Germany}

%

\author{Yuri Gaididei}
\affiliation{Bogolyubov Institute for Theoretical Physics of National Academy of Sciences of Ukraine, 03680 Kyiv, Ukraine}

\date{\today}

%
%
%
%
\begin{abstract}

We show that the presence of a localized curvilinear defect drastically changes magnetic properties of a thin perpendicularly magnetized ferromagnetic film. For a large enough defect amplitude a discrete set of equilibrium magnetization states appears forming a ladder of energy levels. Each equilibrium state has either zero or unit topological charge, i.e. topologically trivial and skyrmion multiplets generally appear. Transitions between the levels with the same topological charge are allowed and can be utilized to encode and switch a bit of information. There is a wide range of geometrical and material parameters, where the skyrmion level has the lowest energy. As a result a periodically arranged curvilinear defects generate a skyrmion lattice as the ground state.
\end{abstract}

\maketitle

\section{Introduction}
An isolated magnetic chiral skyrmion is a localized topologically nontrivial excitation, which may appear in a perpendicularly magnetized ferromagnetic film, when the Dzyaloshinskii-Moriya interaction (DMI) is present. \cite{Nagaosa13,Leonov16,Wiesendanger16}
During the last years isolated skyrmions have been widely considered as data carriers in spintronic data storage and logic devices of a racetrack configuration.\cite{Fert13,Sampaio13,Tomasello14,Zhang15,Krause16,Kang16,Wiesendanger16,Mueller17}
Besides nanotracks\cite{Fert13,Sampaio13,Romming13,Jiang2015,Wiesendanger16} individual skyrmions were obtained in nanodisks.\cite{Sampaio13,Beg15,Buettner15}  Due to nonlocal magnetostatic effects in confined magnetic objects, the skyrmion state can have lower energy as compared to the topologically trivial homogeneous state. \cite{Sampaio13,Beg15}

In contrast to individual skyrmions, their periodic 2D arrays, i.e. skyrmion lattices\cite{Muehlbauer09,Yu10,Yu11,Milde13,Roessler06} are relevant for electronics relying on topological properties of materials. In this regard, dense lattices of small-sized skyrmions facilitate the signal readout in prospective spintronic devices by enhancing the topological Hall effect. \cite{Lee09a,Neubauer09,Kanazawa11,Li13a} Typically, skyrmion lattices are in-field low temperature pocket phases\cite{Muehlbauer09,Yu10,Milde13,Yu11} which hinder their application potential.

\begin{figure*}
	\includegraphics[width=\textwidth]{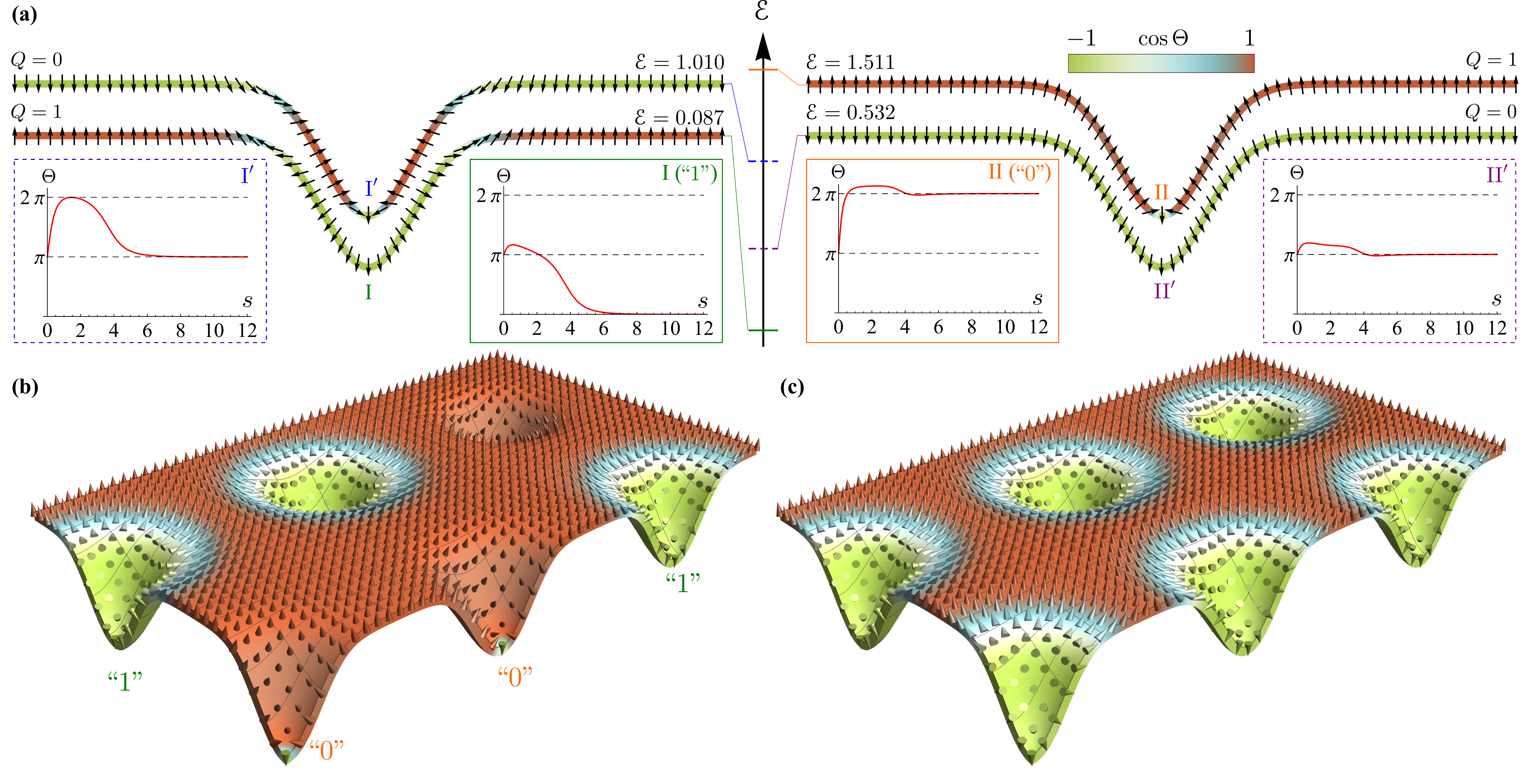}
	\caption{\textbf{Individual skyrmion profiles and skyrmion lattices.} (a): Equilibrium magnetization states of a Gau{\ss}ian concave bump ($\mathcal{A}=-3$, $r_0=1$ and $d=1$) are shown by means of vertical cross-sections. Arrows show the magnetization distribution and color corresponds to the normal component $m_n=\cos\Theta$.  The corresponding solutions $\Theta(s)$ of Eq.~\eqref{eq:Theta} are shown in the insets I, II, I$'$ and II$'$. Vertical axis $\mathcal{E}$ shows distribution of the corresponding energy levels obtained from \eqref{eq:Theta-energy}. (b): Two skyrmion states with big (I) and small (II) radii are shown on the same bumps arranged in a square lattice. These skyrmion solutions can be considered as logical states ``1'' and ``0'' of an information bit. (c) Skyrmion lattice as a ground state.}\label{fig:solutions}
\end{figure*}

Here we demonstrate that magnetic skyrmion can be pinned on a localized curvilinear defect and can have two or more  equilibrium  states with very different skyrmion radius, i.e. one deals with a multiplet of skyrmion states. In this context, a doublet of skyrmion states can be used to represent a single bit of information, see Fig.~\ref{fig:solutions}(b). This unique feature of a skyrmion on a curvilinear defect paves the way towards a new memory concept which is based on immobile skyrmions. 

It is remarkable that when the radii of the skyrmion and the curvilinear defect are comparable, the energy of the skyrmion state can be the lowest one within the class of radially symmetrical solutions. In this way we demonstrate the possibility to realize the lowest energy skyrmion states on a curvilinear defect relying on local interactions only without the need of any magnetic field or magnetostatic effects. As a consequence, a periodically arranged lattice of the defects can generate a skyrmion lattice as a ground state, see Fig.~\ref{fig:solutions}(c). It is important that such a skyrmion lattice exists in zero magnetic field and for a temperature regime, which allows individual skyrmions, e.g. for room temperatures.\cite{Jiang2015,Chen15b} In contrast to the planar case\cite{Heinze11,Kanazawa12} the proposed zero-field lattice does not require four-spin interactions, it can have an arbitrary symmetry and its length scale can be much larger than atomic one. The proposed static reconfigurable lattice of skyrmions opens new exciting perspective for the manipulation and control of spintronic devices relying on the topological Hall effect. \cite{Lee09a,Neubauer09,Kanazawa11,Li13a} Moreover, the feature of the reconfigurability of hte lattice enables new functionality for the skyrmion-based neural network devices. \cite{Huang17a,Prychynenko17,He17}

\section{Model}\label{sec:model}
Similarly to the well studied planar case\cite{Leonov16,Bogdanov89,Bogdanov94,Bogdanov99,Bogdanov01,Komineas15c} the form of a chiral skyrmion is mainly determined by competition of three local interactions: exchange, easy-normal anisotropy and DMI. Thus the energy functional of our model reads 
\begin{equation}\label{eq:E}
E=L\int\left[A \mathscr{E}_\text{ex}+K(1-m_n^2)+D\mathscr{E}_\textsc{d}\right]\mathrm{d}\mathcal{S},
\end{equation}
here $L$ is the film thickness and the integration is performed over the film area. The first term of the integrand is the exchange energy density with $\mathscr{E}_\text{ex}=\sum_{i=x,y,z}(\partial_i\vec{m})^2$, and $A$ being the exchange constant. Here $\vec{m}=\vec{M}/M_s$ is the unit magnetization vector with $M_s$ being the saturation magnetization. The second term is the easy-normal anisotropy where $K>0$ and $m_n=\vec{m}\cdot\vec{n}$ is the normal magnetization component with $\vec{n}$ being the unit normal to the surface. The exchange-anisotropy competition results in the magnetic length $\ell=\sqrt{A/K}$, which determines a length scale of the system. The last term in \eqref{eq:E} represents DMI with $\mathscr{E}_\textsc{d}=m_n\vec\nabla\cdot\vec{m}-\vec{m}\cdot\vec{\nabla}m_n$. Such a kind of DMI originates from the inversion symmetry breaking on the film interface; it is typical for ultrathin films\cite{Crepieux98,Bogdanov01,Thiaville12} or bilayers,\cite{Yang15} and it results in so called N\'{e}el (hedgehog) skyrmions.\cite{Sampaio13,Rohart13} For a surface of rotation with a radially symmetrical magnetization distribution the same type of DMI effectively appears in the exchange term due to curvature effects,\cite{Gaididei14,Sheka15,Kravchuk16a} thus a direct competition takes place. This results in a skyrmion solution of N\'{e}el type. Another types of DMI may lead to a spiral-like skyrmion, which are intermediate ones between N\'{e}el and Bloch types. This case would require a more bulk analysis. 

In our model we disregard nonlocal magnetostatic effects. Still, in stark contrast to the planar case, this is not required for the realization of a skyrmion lowest energy state. \footnote{In other words the magnetostatic contribution is replaced by an effective easy-surface anisotropy, \cite{Slastikov05,Gioia97,Kohn05,Kohn05a} which simply results in a shift of the anisotropy constant $K$.} We also assume magnetization homogeneity along the normal direction, which is valid for $L\lesssim\ell$.

We now consider a curvilinear defect of the film, which is formed by a complete revolution of the curve $\vec{\gamma}=r\vec{e}_x+z(r)\vec{e}_z$
around $z$-axis -- a bump, see Appendix~\ref{app:geom} for details. The parameter $r\ge0$ denotes the distance to the axis of rotation. The surface is assumed to be smooth everywhere, thus $z'(0)=0$. Here and below all distances are considered dimensionless and they are measured in units of the magnetic length $\ell$. Curvilinear properties of the surface at each point are completely determined by two principal curvatures $k_1$ and $k_2$, see the explicit forms in Appendix~\ref{app:geom}.

 The constrain $|\vec{m}|=1$ is utilized by introducing the spherical angular  parameterization $\vec{m}=\sin\theta\cos\phi\vec{e}_s+\sin\theta\sin\phi\vec{e}_\chi+\cos\theta\vec{n}$ in the local orthonormal basis $\{\vec{e}_s,\,\vec{e}_\chi,\,\vec{n}\}$, 
where $\vec{e}_s$ is unit vector tangential to the curve $\vec{\gamma}$, and $\vec{e}_\chi=\vec{n}\times\vec{e}_s$ is the unit vector in azimuthal direction, see Fig.~\ref{fig:surface}. Expressions for $\mathscr{E}_\text{ex}$ and $\mathscr{E}_\textsc{d}$ for a general case of a local curvilinear basis were previously obtained in Ref.~\onlinecite{Gaididei14} and Ref.~\onlinecite{Kravchuk16a}, respectively. Without edge effects (e.g. for a closed surface or for an infinitely large film) the DMI energy density can be reduced to the form 
\begin{equation}\label{eq:Ed}
\mathscr{E}_\textsc{d}=\sin^2\theta\left[2(\vec{\nabla}\theta\cdot\vec{\varepsilon})+\mathcal{H}\right],
\end{equation}
where $\vec{\varepsilon}=\cos\phi\vec{e}_s+\sin\phi\vec{e}_\chi$ is normalized projection of the vector $\vec m$ on the tangential plane and $\mathcal{H}=k_1+k_2$ is the mean curvature. Expression \eqref{eq:Ed} clearly shows the appearance of an effective DMI-driven uniaxial anisotropy proportional to the mean curvature. It has the same curvilinear origin as the recently obtained exchange-driven anisotropy and DMI.\cite{Gaididei14,Sheka15} Depending on sign of the product $D\mathcal{H}$ this anisotropy can be of easy-normal ($D\mathcal{H}>0$) or easy-surface ($D\mathcal{H}<0$) type.\footnote{Note that the sign of each of the quantities $\mathcal{H}$ and $D$ separately is not physically fixed, because it is determined by the chosen direction of the normal vector $\vec n$.}


One can show (see Appendix~\ref{app:eqs-of-motion}) that the total energy \eqref{eq:E} is minimized by a stationary solution $\vec{m}=\sin\Theta\vec{e}_s+\cos\Theta\vec{n}$, where function $\Theta(s)\in \mathbb{R}$ is determined by equation
\begin{equation}\label{eq:Theta}
\begin{split}
\Delta_s\Theta-\sin\Theta\cos\Theta\,\Xi+\frac{r'}{r}(d-2k_2)\sin^2\Theta=\mathcal{H}'.
\end{split}
\end{equation}
Here $s$ is the arc length along $\vec{\gamma}$. The radial part of the Laplace operator reads $\Delta_sf=r^{-1}(rf')'$. Here and everywhere below the prime denotes the derivative with respect to $s$ and the function $r(s)$ completely determines the surface, see Appendix~\ref{app:eqs-of-motion}.
The dimensionless DMI constant $d=D/\sqrt{AK}$ is the only material parameter, which controls the system, and $\Xi=1+r^{-2}r'^2-k_2^2+d\mathcal{H}$. 
It is important to note that any solution of Eq.~\eqref{eq:Theta} and its energy 
\begin{equation}\label{eq:Theta-energy}
\begin{split}
\mathcal{E}=\frac14\int\limits_0^\infty\Bigl[&\Theta'^2+2\Theta'\left(d\sin^2\Theta-k_1\right)+\Xi\sin^2\Theta\\
&-2k_2\frac{r'}{r}\sin\Theta\cos\Theta+k_1^2+k_2^2\Bigr]r(s)\mathrm{d}s
\end{split}
\end{equation}
are invariant with respect to the transformation $\Theta\to\Theta+\pi$, i.e. any solution is doubly degenerate with respect to the replacement $\vec{m}\to-\vec{m}$.\footnote{If solution $\vec{m}$ is stable then solution $-\vec{m}$ is also stable, see Appendix~\ref{sec:stab}.} Consequently, one can fix the boundary condition $\Theta(0)=\pi$ at the bump center without loss of generality and consider different boundary conditions at the infinity: $\Theta(\infty)=n\pi$ with $n\in\mathbb{Z}$. In \eqref{eq:Theta-energy} and everywhere below we use the normalized energy $\mathcal{E}=E/E_\textsc{bp}$, where  $E_\textsc{bp}=8\pi AL$ is energy of the Belavin-Polyakov soliton.\cite{Belavin75} The same invariance takes place for the transformation $k_1\to-k_1$, $k_2\to-k_2$, $d\to-d$, $\Theta\to2\pi-\Theta$. This property is reflected in the symmetry of the diagram of skyrmion states, see Fig.~\ref{fig:diagram}.

Following Ref.~\onlinecite{Kravchuk16a} one can show that topological charge (mapping degree to $S^2$) of such a radially symmetrical solution on a localized bump reads $Q=\frac12[\cos\Theta(\infty)-\cos\Theta(0)]$, see Appendix~\ref{app:Q} for details. It means that only values $Q=0$ (for odd $n$) or $Q=1$ (for even $n$) are possible. A state with $Q=-1$ appears under the transformation $\vec{m}\to-\vec{m}$ applied to the state with $Q=1$. 

Due to the presence of the right-hand-part driving term in Eq.~\eqref{eq:Theta} the trivial solutions $\Theta\equiv0,\pi$ (i.e. $\vec{m}=\pm\vec{n}$) are generally not possible. It means that even for large anisotropy the magnetization vector deviates from the normal direction, except surfaces with $\mathcal{H}=\text{const}$, e.g. planar films, spherical and minimal surfaces. Such a prediction was previously made in Ref.~\onlinecite{Gaididei14}. An analogous driving appears in 1D curvilinear wires and results in curvature induced domain wall motion along the curvature gradient.\cite{Yershov15b} Thus, Eq.~\eqref{eq:Theta} makes one expect a leading role of the mean curvature gradient in the analogous curvature induced skyrmion motion. 

\begin{figure}
	\includegraphics[width=\columnwidth]{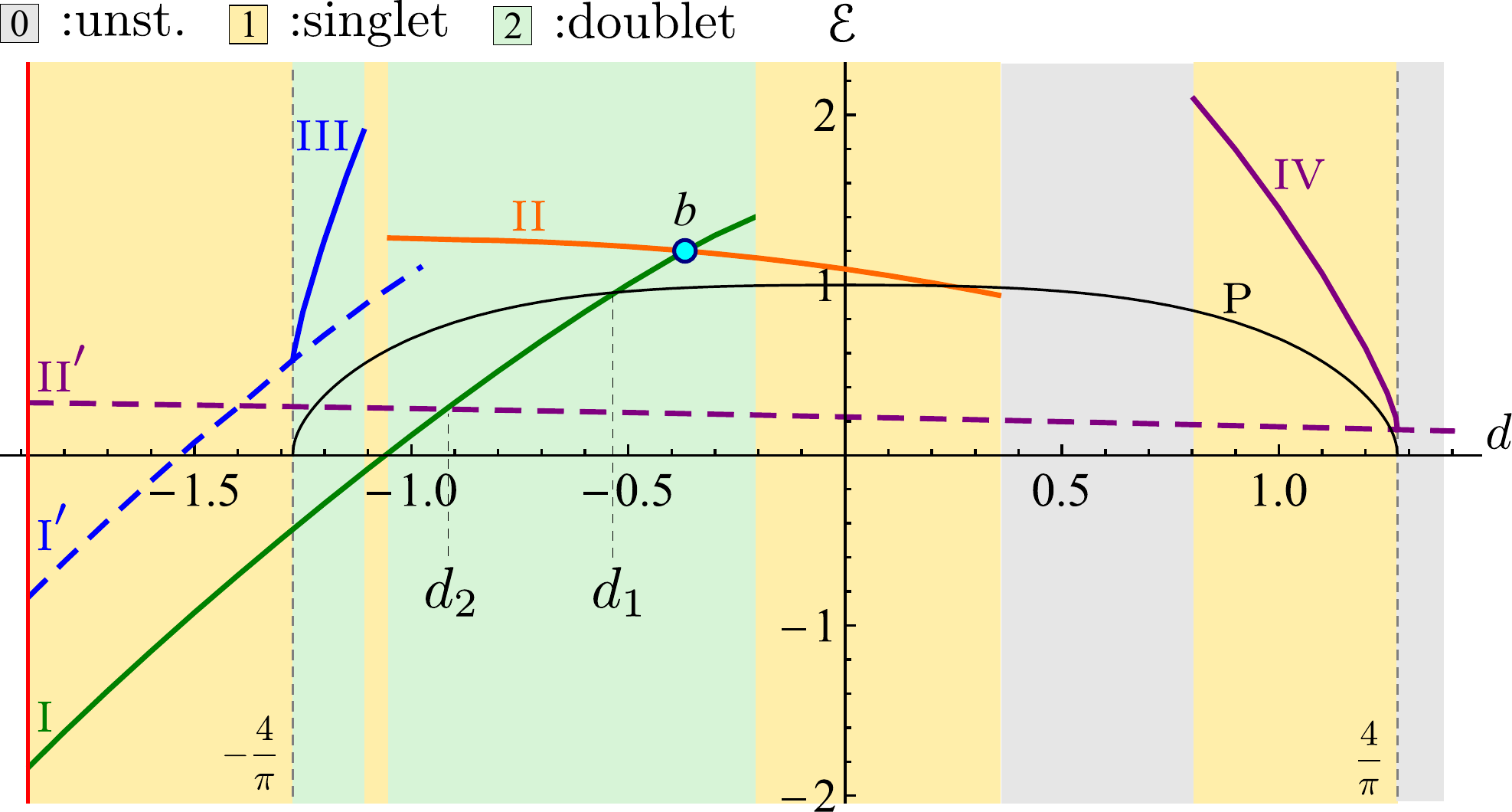}
	\caption{\textbf{Energies of different solutions.} Solid lines I-IV and dashed lines I$'$, II$'$ show energies \eqref{eq:Theta-energy} of topological non-trivial (skyrmion) and trivial states, respectively for the bump with $\mathcal{A}=2$ and $r_0=1$. Energy of the planar skyrmion is shown by the thin line P. States I, II, I$'$, II$'$ are similar to the same name states in Fig.~\ref{fig:solutions}. States III and IV correspond to skyrmions whose radius much exceeds the lateral bump size, see Figs.~\ref{fig:spectrum-a},~\ref{fig:spectrum-d}. The background filling corresponds to the number of stable skyrmion states, see also Fig.~\ref{fig:diagram}. }\label{fig:energies}
\end{figure}

In the planar film limit $k_1=k_2\equiv0$, $\mathcal{H}\equiv0$ and $r(s)=s$. In this case Eq.~\eqref{eq:Theta} is transformed into the well-known\cite{Bogdanov94,Rohart13,Komineas15c,Leonov16} chiral skyrmion equation. Such a planar system is controlled by the only parameter $d$. There is the critical value $d_0=4/\pi$, which separates two ground states, namely the uniform state $\vec{m}=\vec{n}$ for the case $|d|<d_0$, and helical periodical state for $|d|>d_0$.\cite{Bogdanov94,Rohart13,Komineas15c,Leonov16} For the case  $|d|<d_0$ the planar form of Eq.~\eqref{eq:Theta} has a stable topological ($Q=1$) solution -- a skyrmion, which has the following features: (i) for a given value of $d$ the skyrmion solution is unique; (ii) the skyrmion energy is always higher than energy of the uniform perpendicular state, i.e. the planar skyrmion is an excitation of the ground state. As we show below, these well-known properties are violated in the general case of the curvilinear defect.

\section{Gau{\ss}ian bump}
As an example, we consider a class of localized curvilinear defects in form $z(r)=\mathcal{A}e^{-r^2/(2r_0^2)}$.
Here amplitudes $\mathcal{A}>0$ and $\mathcal{A}<0$ correspond to bumps that are convex or concave, respectively, and $r_0$ determines the bump width. In Fig.~\ref{fig:solutions}(a) we demonstrate stable equilibrium solutions of Eq.~\eqref{eq:Theta} for certain values of parameters. There is a number of principal differences as compared to the planar case: \\
(i)  Topological ($Q=1$) as well as trivial ($Q=0$) solutions are generally not unique: for given values of geometrical and material parameters a set of equilibrium magnetization states can appear with a ladder of energy levels. This makes the curvilinear defect conceptually similar to a quantum well with a finite number of discrete energy levels. However, in contrast to the quantum systems the transitions between levels with the same $Q$ are only allowed. Such a transitions are expected to be accompanied by emission or absorption of magnons.\\
(ii) The lowest energy level can be topological non-trivial ($Q=1$). Recently it was shown\cite{Sampaio13} that due to nonlocal magnetosatic effects the skyrmion state of a finite size disk can have lower energy as compared to the topologically trivial homogeneous state. Here we thus demonstrate that the same is possible in curvilinear systems due to local interactions only. As a consequence, curvilinear defects arranged in a periodical lattice generate a zero-field skyrmion lattice\cite{Heinze11,Kanazawa12} as a ground state of the system, see Fig.~\ref{fig:solutions}(c). 

\begin{figure}
	\includegraphics[width=\columnwidth]{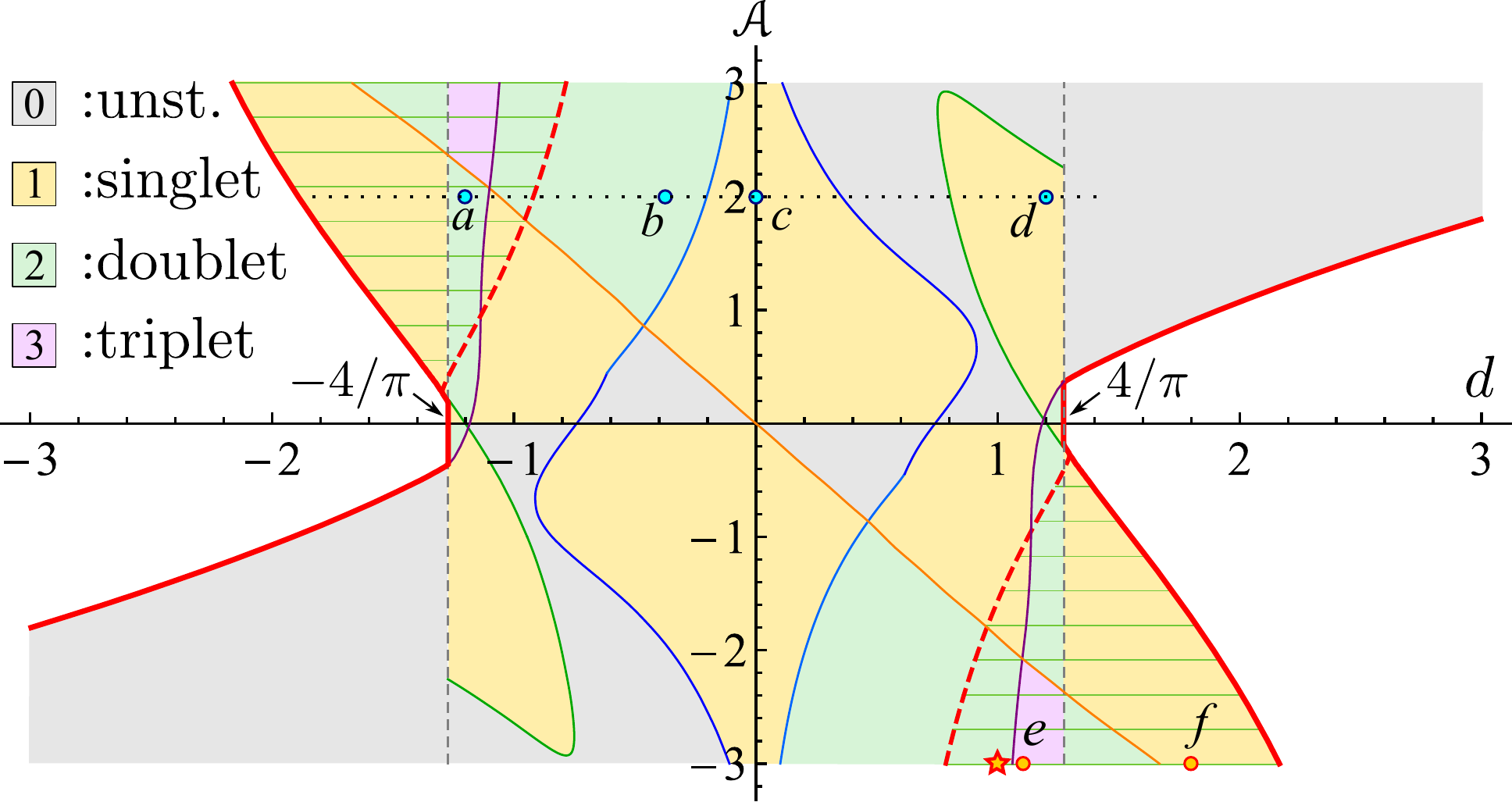}
	\caption{\textbf{Diagram of skyrmion states for Gau{\ss}ian bump with $r_0=1$.} In the white area the skyrmion solutions does not exist. Number of any other area (see legend) coincides with the number of stable skyrmion solutions. At least one skyrmion solution exists within the gray area `0', however the bump center is a position of unstable equilibrium for it. Within the other areas the corresponding number of skyrmion are pinned at the bump center. The horizontal dashing shows areas where the lowest energy level is skyrmion one. Star marker shows parameters of Fig.~\ref{fig:solutions}. The solutions spectra for points \textit{a-f} are presented in Appendix~\ref{sec:solutions}. Dotted horizontal line $\mathcal{A}=2$ corresponds to Fig.~\ref{fig:energies}. }\label{fig:diagram}
\end{figure}

Let us consider skyrmions of small and big radii, which are shown in the Fig.~\ref{fig:solutions} as states I(``1'') and II(``0''), respectively. Their radii\footnote{Skyrmion radius $R$ is determined as solution of equation $\cos\Theta(R)=0$. For the parameters of Fig.~\ref{fig:solutions} the small and big skyrmion radii are $R_\text{sml}\approx0.14$ and $R_\text{big}\approx3.7$, respectively; $\mathcal{K}(s)$ has extreme points $s_1=0$ and $s_2\approx3.3$.} are close to extrema points of the Gau{\ss} curvature $\mathcal{K}=k_1k_2$, which plays an important role in a coupling between topological defects and curvature.\cite{Vitelli04,Turner10} Hence, one can expect an important role of the curvature in the skyrmions stabilization. On the other hand, the radius of skyrmion II is of one order of magnitude smaller than radius of the skyrmion, which is stabilized by the intrinsic DMI in a planar film for the same value of $d$. Thus, the small radius skyrmion is stabilized mostly by the curvature effects,\cite{Kravchuk16a,Carvalho-Santos15a,Carvalho-Santos13,VilasBoas15} while the big radius skyrmion is stabilized due to the simultanious action of the intrinsic DMI and curvature. Structures similar to the big radius skyrmions were previously observed experimentally in Co/Pd and Co/Pt multilayer films containing an array of curvilinear defects in form of spherical concavities\cite{Makarov07,Makarov09a} as well as convexes.\cite{Brombacher09,Streubel16a} The topologically trivial state I$'$ can be treated as a joint state of small and big radii skyrmions, which compensate topological charges of each other. And the state II$'$ is an intermediate one between uniform $\vec{m}=-\vec{e}_z$ and normal $\vec{m}=-\vec{n}$ states, what reflects the competition between exchange and anisotropy interactions. Note that states I and I$'$ as well as states II and II$'$ differ in presence or absence of the small-radius skyrmion at the bump center. In Fig.~\ref{fig:solutions}(a) we show only stable solutions with $\Delta\Theta=|\Theta(\infty)-\Theta(0)|\le\pi$. Solutions with the larger phase incursion, so called skyrmioniums\cite{Komineas15c,Finazzi13} or target skyrmions,\cite{Bogdanov99,Liu15,Rohart13,Beg15,Leonov14a} are in principle also possible.

The appearance of skyrmions of type I (big radius) and type II (small radius) is a common feature of the considered curvilinear defects, and takes place for concave as well as for convex geometries. In order to illustrate the last statement we show the energies $\mathcal{E}(d)$ for all equilibrium states, which appear for a convex bump, see Fig.~\ref{fig:energies}. For the given geometrical parameters we found numerically all solutions of Eq.~\eqref{eq:Theta} with $\Delta\Theta\le\pi$ for each value $d$. Then a stability analysis (see Appendix~\ref{sec:stab}) was applied for each of the solutions. Finally, four stable topological (skyrmion) solutions (lines I-IV) and two stable non-topological solutions (lines I$'$ and II$'$) are found. The magnetization distributions, that correspond to all of these solutions, are shown in Appendix~\ref{sec:solutions}. Lines I and II correspond to the considered above big (``1'') and small (``0'') radius skyrmions, respectively. Remarkably these states can have equal energies -- point \textit{b} in Fig.~\ref{fig:energies}. This makes the proposed application for the storing of a bit of information more practically relevant: switching between states ``0'' and ``1'' can be easily controlled by application of pulse of magnetic field directed along or against the vertical axis.

As well as for the concave geometry (Fig.~\ref{fig:solutions}) the big radius skyrmion on a convex bump can have the lowest energy in the system (the range $d<d_2$). It is important to note that there is a range of parameters $-4/\pi<d<d_1$ where a skyrmion on a bump has lower energy than a planar skyrmion for the same $d$. This implies that flexible enough planar films can spontaneously undergo a skyrmion induced deformation. Such a soliton-induced magnetic film deformation was earlier predicted for cylindrical geometries. \cite{Dandoloff95,Villain-Guillot95,Saxena97,Saxena98,Saxena98a}


In order to systematize possible skyrmion solutions, that can appear on Gau{\ss}ian bumps, we build a diagram of skyrmion states, see Fig.~\ref{fig:diagram}. We apply the same method as for the case of Fig.~\ref{fig:energies}, but restricting ourselves with skyrmion solutions. The following general features can be established: (i) The range of skyrmions existence widens with increasing of the bump amplitude. (ii) For a wide range of parameters (gray area `0') the skyrmion centered on the bump experiences a displacement instability because the bump center is a position of unstable equilibrium.
(iii) In the vicinity of the critical value $d=\pm4/\pi$ there is a wide area of parameters (the dashed area), where the skyrmion state has the lowest energy in the class of radially symmetrical solutions. If such a skyrmion is the ground state of the system, then one can consider the described bump as a generator of skyrmions. Indeed, a spontaneous formation of the skyrmion is expected due to thermal fluctuations. Applying now a strong enough in-surface spin-polarized current can move the skyrmion from the bump into the planar part of the film. This operation can be then repeated.

\section{Conclusions}
We have generalized the skyrmion equation for the case of an arbitrary surface of rotation. Considering specifically a Gau{\ss}ian bump we have shown that its skyrmion solution is generally not unique --- a discrete ladder of equilibrium skyrmion states appears. We propose to use a suitably shaped curvilinear defect with a doubly degenerate skyrmion state as carrier of a bit of information. We also predict the effect of spontaneous deformation of an elastic magnetic film under skyrmion influence. Finally, we found a wide range of parameters, where a skyrmion pinned on the bump has lower energy then other possible states. This feature can be used for generating of a ground state zero-field skyrmion lattice.

\section{Acknowledgments}
V.P.K. acknowledges the Alexander von Humboldt Foundation for the support. This work has been supported by the DFG via SFB 1143 and by ERC within the EU 7th Framework Programme (ERC Grant No. 306277) and the EU FET Programme (FET-Open Grant No. 618083). We acknowledge Prof. Avadh Saxena for fruitful discussions.


\appendix

\section{Geometrical properties of a surface of revolution}\label{app:geom}

\begin{figure}
	\includegraphics[width=\columnwidth]{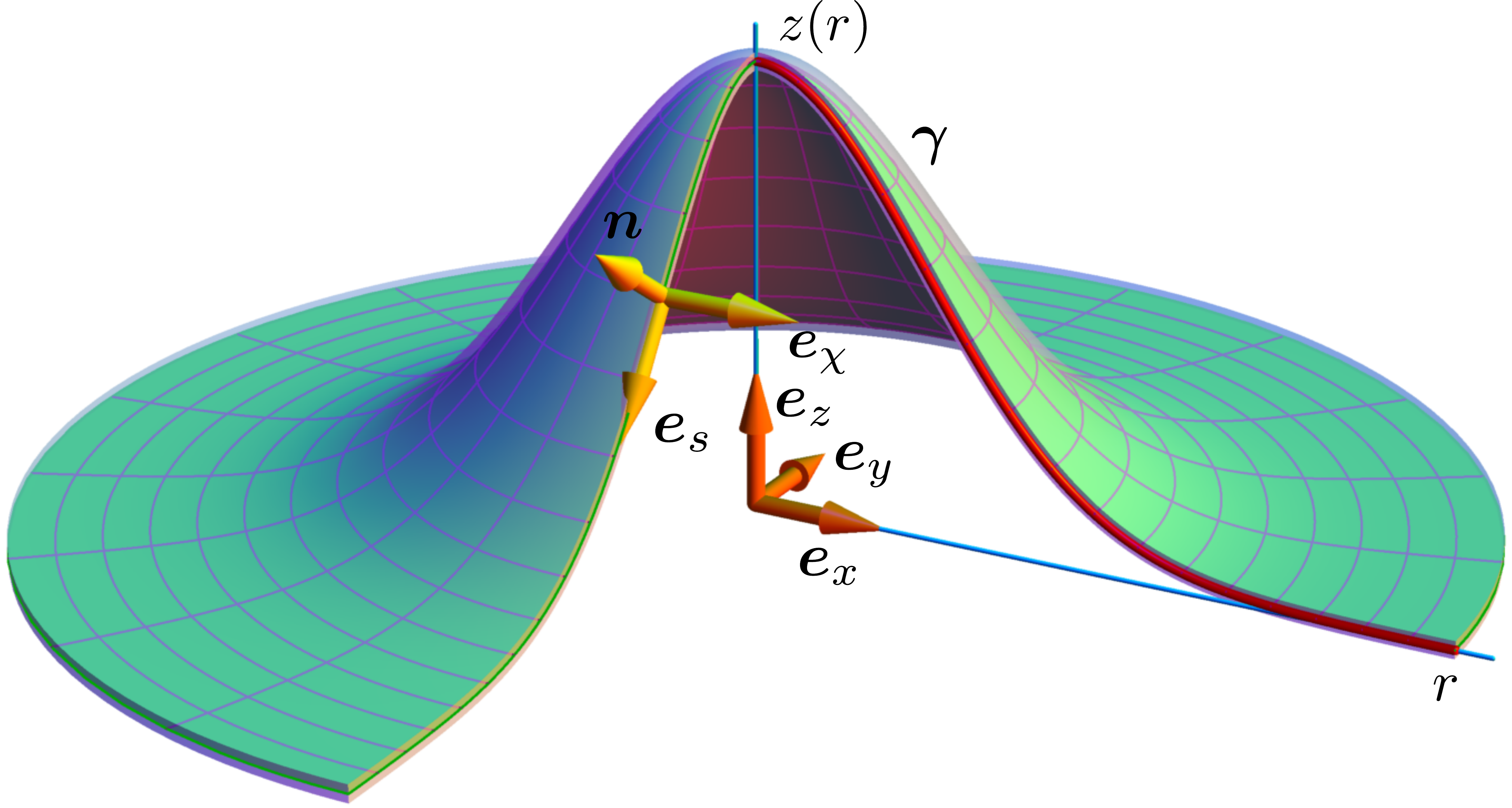}
	\caption{\textbf{Details of the geometry}. A curvilinear defect is considered as a surface of rotation, which is formed by a complete revolution of the curve $\vec{\gamma}=r\vec{e}_x+z(r)\vec{e}_z$ around $z$-axis. Cartesian $\{\vec{e}_x,\vec{e}_y,\vec{e}_z\}$ and curvilinear $\{\vec{e}_s,\vec{e}_\chi,\vec{n}\}$ frames of reference are introduced.}\label{fig:surface}
\end{figure}
Curvilinear properties of the considered surface of rotation (see Fig.~\ref{fig:surface}) are completely determined by two principal curvatures 
\begin{equation}\label{eq:k1-k2}
k_1(r)=\frac{z''(r)}{[1+z'(r)^2]^{3/2}},\quad k_2(r)=\frac{z'(r)}{r\sqrt{1+z'(r)^2}}.
\end{equation}
Here $k_1$ is curvature of ${\vec{\gamma}}$ while  $k_2$ is curvature of the curve created by the section of the surface with the plane perpendicular to $\vec{\gamma}$ at each given point. For a surface of rotation the principal curvatures are not independent: for the known $k_1$ the curvature $k_2$ can be found as a solution of differential equation $rk_2'(r)+k_2(r)=k_1(r)$ with the initial condition $k_2(0)=k_1(0)$. Principal curvatures are invariants of the surface, so they can be considered as functions of the point. 

\begin{figure}
	\includegraphics[width=0.7\columnwidth]{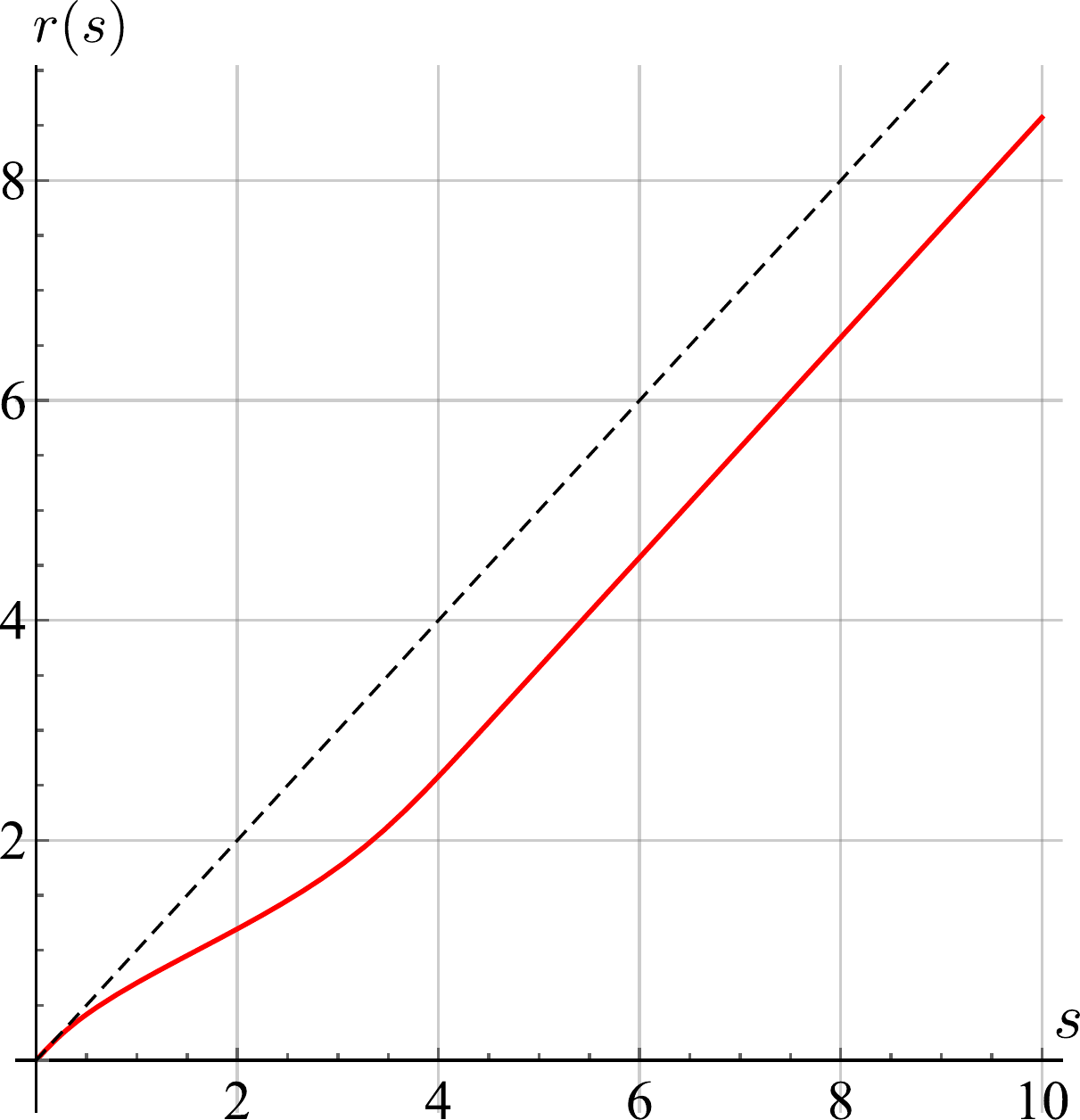}
	\caption{\textbf{Arc length parameterization.} Dependence $r(s)$ for the Gau{\ss}ian bump $z(r)=\mathcal{A}e^{-r^2/(2r_0^2)}$ with $\mathcal{A}=3$ and $r_0=1$.}\label{fig:rvss}
\end{figure}

In the following it is convenient to use the natural parameter $s$ (the arc length along $\vec{\gamma}$) instead of parameter $r$. The functional dependence $r=r(s)$ between the parameters is determined by differential equation $r'(s)\sqrt{1+z'_r(r(s))^2}=1$ supplemented with the initial condition $r(0)=0$. An example of the dependence $r(s)$ for a Gau{\ss}ian bump is shown in Fig.~\ref{fig:rvss}.
Function $r(s)$ completely determines curvilinear properties of the surface:
\begin{equation}\label{eq:k1_k2_s}
k_1(s)=\frac{r''}{\sqrt{1-r'^2}},\quad k_2(s)=-\frac{\sqrt{1-r'^2}}{r}.
\end{equation}
From \eqref{eq:k1_k2_s} one obtains the following useful properties
\begin{equation}\label{eq:k1k2}
\mathcal{K}=-r''/r,\qquad [k_1(s)-k_2(s)]r'=k_2'(s)r.
\end{equation}

\section{Equilibrium solutions}\label{app:eqs-of-motion}

Equilibrium solutions are determined by set of equations 
\begin{equation}\label{eq:equilib}
\frac{\delta E}{\delta\theta}=0,\qquad \frac{\delta E}{\delta\phi}=0,
\end{equation}
where the energy functional $E$ is defined in \eqref{eq:E}.
Exchange energy density for an arbitrary curvilinear film reads\cite{Gaididei14,Sheka15}
\begin{equation}\label{eq:Eex}
\mathscr{E}_\text{ex}=\left[\vec\nabla\theta-\vec{\Gamma}\right]^2+\left[\sin\theta(\vec{\nabla}\phi-\vec{\Omega})-\cos\theta\partial_\phi\vec{\Gamma}\right]^2.
\end{equation}
Here $\vec{\Omega}$ is vector of spin connection and $\vec{\Gamma}=||h_{\alpha\beta}||\cdot\vec{\varepsilon}$, where $||h_{\alpha\beta}||$ is Weingarten map. For details see Refs.~\onlinecite{Kamien02,Bowick09,Kravchuk16a}. Substituting into the total energy \eqref{eq:E} the contributions from the exchange energy \eqref{eq:Eex}, DMI $\mathscr{E}_\textsc{d}$ together with the easy-normal anisotropy contribution one can write equations \eqref{eq:equilib} in the following expanded form
	\begin{equation}\label{eq:general}
	\begin{split}
	&\Delta\theta-\sin\theta\cos\theta\left[1+(\vec{\nabla}\phi-\vec{\Omega})^2-(\partial_\phi\vec{\Gamma})^2+d\mathcal{H}\right]\\
	&-\vec{\nabla}\!\cdot\!\vec{\Gamma}+\cos2\theta(\vec{\nabla}\phi-\vec{\Omega})\!\cdot\!\partial_\phi\vec{\Gamma}+d\sin^2\theta\vec{\nabla}\!\cdot\!\vec{\varepsilon}=0,\\
	&\vec{\nabla}\!\cdot\!\left[\sin^2\theta(\vec{\nabla}\phi-\vec{\Omega})\right]-\sin\theta\cos\theta\left[(\vec{\nabla}\phi-\vec{\Omega})\!\cdot\!\vec{\Gamma}+\vec{\nabla}\!\cdot\!\partial_\phi\vec{\Gamma}\right]\\
	&+\sin^2\theta\left[2(\vec{\nabla}\theta\cdot\partial_\phi\vec{\Gamma})-\vec{\Gamma}\!\cdot\!\partial_\phi\vec{\Gamma}-d(\vec{\nabla}\theta\!\cdot\!\vec{\varepsilon})\right]=0.
	\end{split}
	\end{equation}
For the case of the considered surface of rotation the vector of spin connection is $\vec{\Omega}=-\vec{e}_\chi r'/r$, and $\vec{\Gamma}=k_1\cos\phi\vec{e}_s+k_2\sin\phi\vec{e}_\chi$. In this case one can see that Eqs.~\eqref{eq:general}
have solution $\phi=\Phi=0,\,\pi$, and $\theta=\Theta(s)$, wherein the second equation in \eqref{eq:general} becomes an identity while the first one obtains the form
\begin{equation}
\begin{split}
\Delta_s\Theta-\sin\Theta\cos\Theta\,\Xi+\mathcal{C}\frac{r'}{r}(d-2k_2)\sin^2\Theta=\mathcal{C}\mathcal{H}'.
\end{split}
\end{equation}
Constant $\mathcal{C}=\cos\Phi=\pm1$ determines sign of the radial component $m_s=\mathcal{C}\sin\Theta$, while the azimuthal component $m_\chi=0$ is absent. In this case without loss of generality one can fix the value $\mathcal{C}=+1$ by expanding the range of values of function $\Theta(s)$ to the full circumference $\Theta\in\mathbb{R}$.

 Note, that such a simple solution is possible for the interfacing DMI, while for another types of DMI the set of equations \eqref{eq:general} can not be generally uncoupled. 
 
\section{Topological charge calculation}\label{app:Q}
Let us calculate topological charge $Q$ -- degree of the  mapping to sphere $S^2$:
\begin{equation}\label{eq:Q}
Q=\frac{1}{4\pi}\int\mathcal{J}\,\text{d}\mathcal{S}.
\end{equation}
Here the mapping Jacobian $\mathcal{J}$ reads\cite{Kravchuk16a}
\begin{equation}\label{eq:J}
\begin{split}
\mathcal{J}=&-\sin\theta\,\vec{n}\cdot\left[(\vec{\nabla}\theta-\vec{\Gamma})\times(\vec{\nabla}\phi-\vec{\Omega})\right]\\
&+\cos\theta\,\vec{n}\cdot\left[\vec{\nabla}\theta\times\partial_\phi\vec{\Gamma}\right]-\cos\theta\,\mathcal{K},
\end{split}
\end{equation}
where $\vec{\Gamma}$ and $\vec{\Omega}$ are defined in the previous section. For the case of solution $\theta=\Theta(s)$ and $\phi=0,\pi$ determined  on the considered surface of rotation one has $\vec{\nabla}\theta=\vec{e}_s\Theta'$, $\vec{\nabla}\phi=\vec{0}$, $\vec{\Gamma}=\mathcal{C}k_1\vec{e}_s$, $\partial_\phi\vec{\Gamma}=\mathcal{C}k_2\vec{e}_\chi$ and $\vec{\Omega}=-\vec{e}_\chi r'/r$. The substitution into \eqref{eq:J} with the subsequent integration \eqref{eq:Q} results in
\begin{equation}\label{eq:Q-res}
Q=\left.\frac{1}{2}\left[r'(s)\cos\Theta(s)\right]\right|_0^\infty=\left.\frac{1}{2}\frac{\cos\Theta(r)}{\sqrt{1+z'(r)^2}}\right|_0^\infty.
\end{equation}
When integrating \eqref{eq:Q} one should use the properties \eqref{eq:k1k2}.
For a smooth and localized curvilinear defect one has $z'(0)=z'(\infty)=0$, thus \eqref{eq:Q-res} result in the common formula
\begin{equation}\label{eq:Q-final}
Q=\frac12\left[\cos\Theta(\infty)-\cos\Theta(0)\right].
\end{equation}

\section{Stability analysis}\label{sec:stab}
Here we consider stability of the obtained stationary solutions $\Theta$ and $\Phi$ with respect to infinitesimal deviations:
\begin{equation}\label{eq:deviations}
\theta=\Theta+\vartheta,\qquad \phi=\Phi+\varphi/\sin\Theta,
\end{equation} 
where $|\vartheta|\ll1$ and $|\varphi|\ll1$. In this case the total normalized energy \eqref{eq:E} can be presented in form $\mathcal{E}\approx \mathcal{E}_0+\epsilon$, where $\mathcal{E}_0=\mathcal{E}[\Theta,\Phi]$ is energy of the unperturbed state, and harmonic part of the energy increase reads
\begin{equation}\label{eq:E-harm}
\epsilon=\frac{1}{8\pi}\int\vec{\Psi}^\textsc{t}\hat{\mathrm{H}}\vec{\Psi}\,\text{d}\mathcal{S}.
\end{equation}
Here $\vec{\Psi}=(\vartheta,\varphi)^\textsc{t}$ and
\begin{equation}\label{eq:H}
\hat{\mathrm{H}}=\begin{pmatrix}
-\Delta+U_1&W\partial_\chi\\
-W\partial_\chi&-\Delta+U_2
\end{pmatrix}.
\end{equation}
The Laplace operator has the form $\Delta=\Delta_s+r^{-2}\partial_{\chi\chi}^2$ and potentials are as follows
\begin{equation}\label{eq:UW}
\begin{split}
U_1=&\cos2\Theta\,\Xi-\mathcal{C}\frac{r'}{r}(d-2k_2)\sin2\Theta,\\
U_2=&\cos^2\Theta\,\Xi-\Theta'^2+k_2^2-k_1^2-\mathcal{C}(d-2k_1)\Theta'\\
&-\mathcal{C}\frac{r'}{r}(d-2k_2)\sin\Theta\cos\Theta,\\
W=&2\frac{r'}{r^2}\cos\Theta-\frac{\mathcal{C}}{r}(d-2k_2)\sin\Theta.
\end{split}
\end{equation}
Using the Fourier series expansion
\begin{equation}\label{eq:Fourier}
\begin{split}
&\vartheta(s,\chi)=\frac{f_0(s)}{2}+\sum\limits_{\mu\ne0}\left[f_\mu(s)\cos\mu\chi+\bar{f}_\mu(s)\sin\mu\chi\right],\\
&\varphi(s,\chi)=\frac{\bar{g}_0(s)}{2}+\sum\limits_{\mu\ne0}\left[\bar{g}_\mu(s)\cos\mu\chi+g_\mu(s)\sin\mu\chi\right]
\end{split}
\end{equation}
one can present energy \eqref{eq:E-harm} in form
\begin{equation}\label{eq:epsilon}
\epsilon=\frac18\sum\limits_{\mu}\int\!r(s)\left[\vec{\psi}_\mu^\textsc{t}\hat{\mathscr{H}}_\mu\vec{\psi}_\mu+\bar{\vec{\psi}}_\mu^\textsc{t}\hat{\mathscr{H}}_{-\mu}\bar{\vec{\psi}}_\mu\right]\text{d}s,
\end{equation}
where $\mu\in\mathbb{Z}$, $\vec{\psi}_\mu=(f_\mu,g_\mu)^\textsc{t}$, $\bar{\vec{\psi}}_\mu=(\bar{f}_\mu,\bar{g}_\mu)^\textsc{t}$ and 
\begin{equation}\label{eq:Hm}
\hat{\mathscr{H}}_\mu=\begin{pmatrix}
-\Delta_s+\frac{\mu^2}{r^2}+U_1&\mu W\\
\mu W&-\Delta_s+\frac{\mu^2}{r^2}+U_2
\end{pmatrix}.
\end{equation}
Stationary solution ($\Theta$, $\Phi$) corresponds to a minimum of energy iff $\epsilon>0$ for any $\vec{\psi}_\mu$ and $\bar{\vec{\psi}}_\mu$. Taking into account that eigenvalues $\lambda_i^{(\mu)}$ of operator $\hat{\mathscr{H}}_\mu$ do not depend on sign of $\mu$, one can state that positiveness of all $\lambda_i^{(\mu)}$ for all is a sufficient condition for the energy minimum, i.e. stability.

In order to check stability of a solution ($\Theta$, $\Phi$) we found numerically the lowest eigenvalues $\Lambda^{(\mu)}=\min_i\lambda_i^{(\mu)}$ of operators $\hat{\mathscr{H}}_\mu$, with $\mu=0,1,\dots4$. If all $\Lambda^{(\mu)}>0$, we conclude that the solution ($\Theta$, $\Phi$) is stable. In case of the skyrmion solution one can say that the bump effectively generates pinning potential. If $\Lambda^{(1)}<0$ and $\Lambda^{(\mu\ne1)}>0$, then we conclude that the bump effectively generates repulsing potential. One can say about displacement instability in this case. 
For the cases $\Lambda^{(0)}<0$ and $\Lambda^{(2)}<0$ one can say about radial\cite{Bogdanov99} and elliptical\cite{Bogdanov94a,Bogdanov99} instability, respectively.

\section{Examples of possible equilibrium magnetization states}\label{sec:solutions}

Here we present spectra of stable solutions of Eq.~\eqref{eq:Theta}, which correspond to the points \textit{a--f} in Fig.~\ref{fig:diagram}.

\begin{figure*}[h]
	\includegraphics[width=\textwidth]{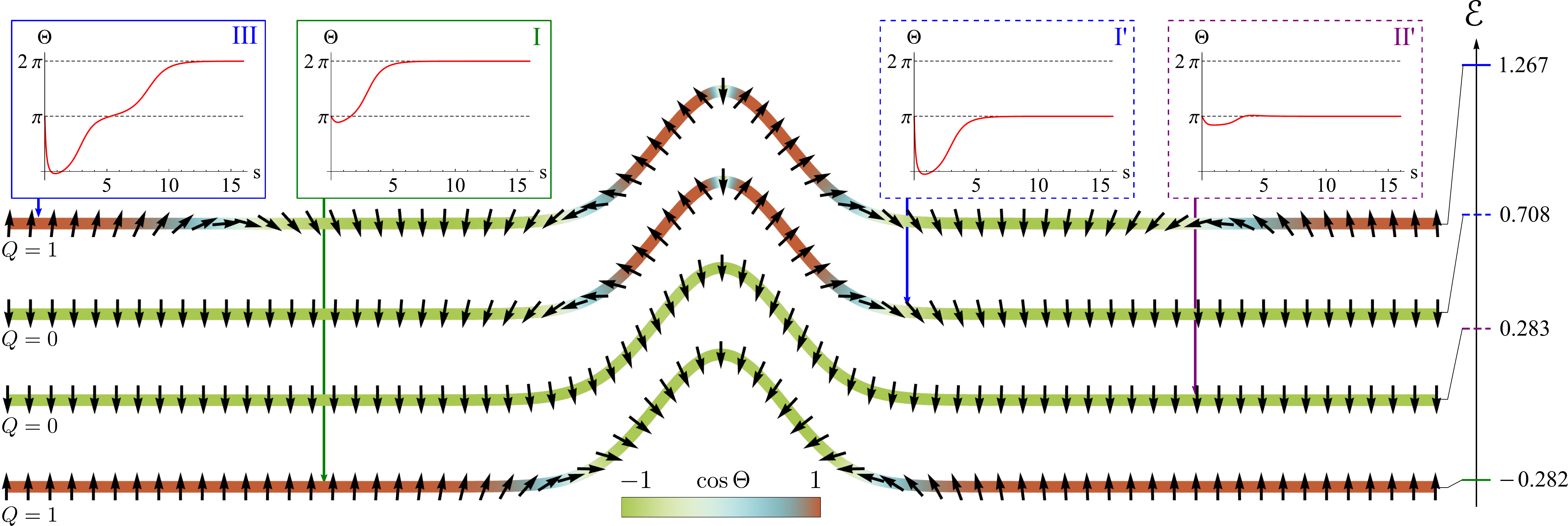}
	\caption{Spectrum of equilibrium magnetization states, which correspond to point ``$a$'' in Fig.~\ref{fig:diagram} --- $\mathcal{A}=2$, $r_0=1$, $d=-1.2$. The lowest I and the highest III energy skyrmions correspond to lines I and III in Fig.~\ref{fig:energies}, respectively. Non-topological solutions  I$'$ and II$'$  correspond to the lines I$'$ and II$'$, respectively.}\label{fig:spectrum-a}
\end{figure*}

\begin{figure*}[h]
	\includegraphics[width=\textwidth]{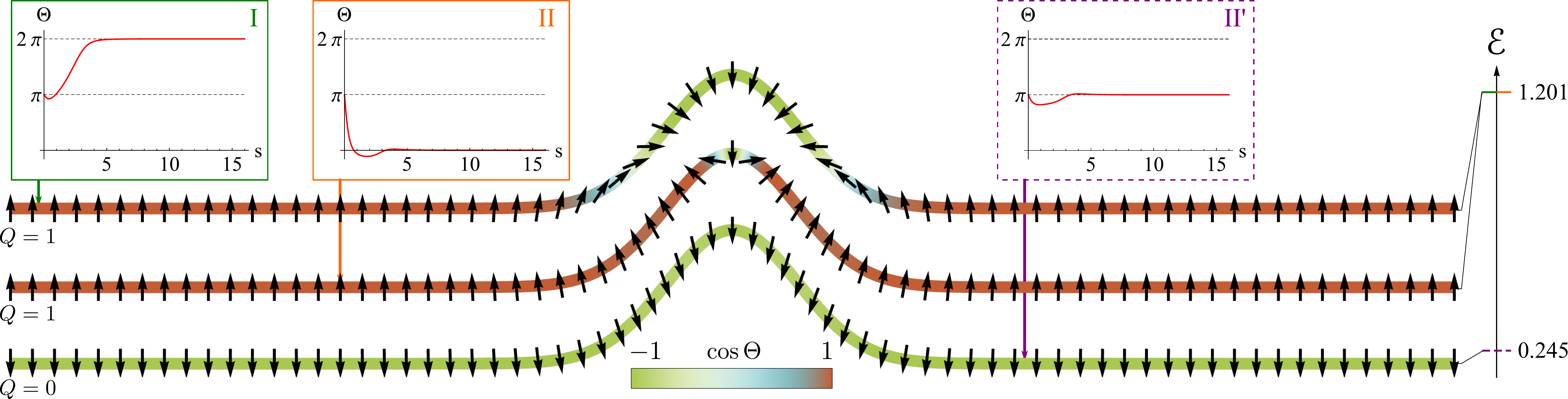}
	\caption{Spectrum of equilibrium magnetization states, which correspond to point ``$b$'' in Figs.~\ref{fig:diagram}, \ref{fig:energies} --- $\mathcal{A}=2$, $r_0=1$, $d=-0.368$. The big I and small II radius skyrmions correspond to lines I and II in Fig.~\ref{fig:energies}, respectively. Non-topological solution II$'$ corresponds to the line II$'$}\label{fig:spectrum-b}
\end{figure*}

\begin{figure*}[h]
	\includegraphics[width=\textwidth]{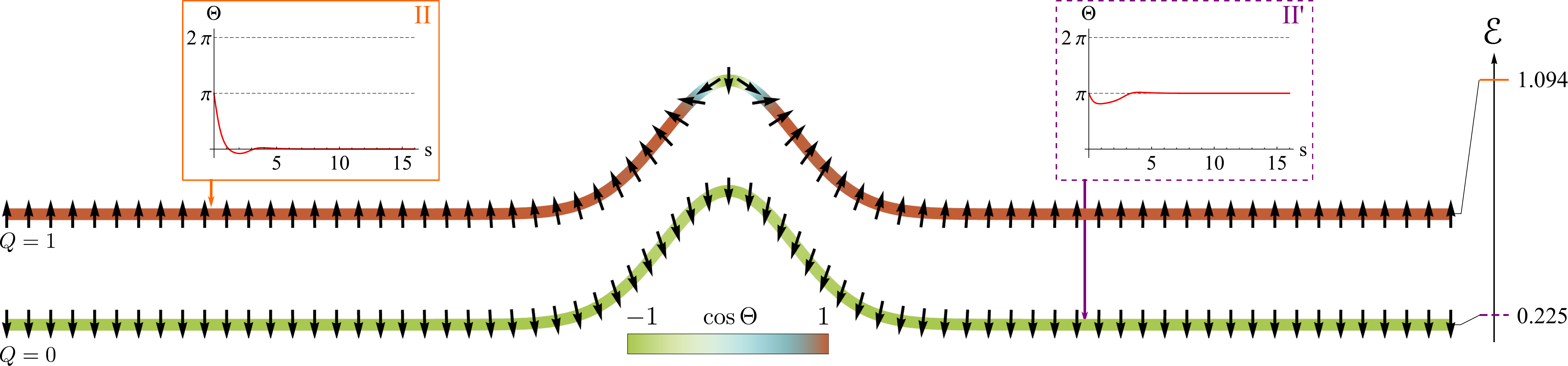}
	\caption{Spectrum of equilibrium magnetization states, which correspond to point ``$c$'' in Fig.~\ref{fig:diagram} --- $\mathcal{A}=2$, $r_0=1$, $d=0$. The curvature stabilized skyrmion II and non-topological solution II$'$ correspond to lines II and II$'$ in Fig.~\ref{fig:energies}, respectively.}\label{fig:spectrum-c}
\end{figure*}

\begin{figure*}[h]
	\includegraphics[width=\textwidth]{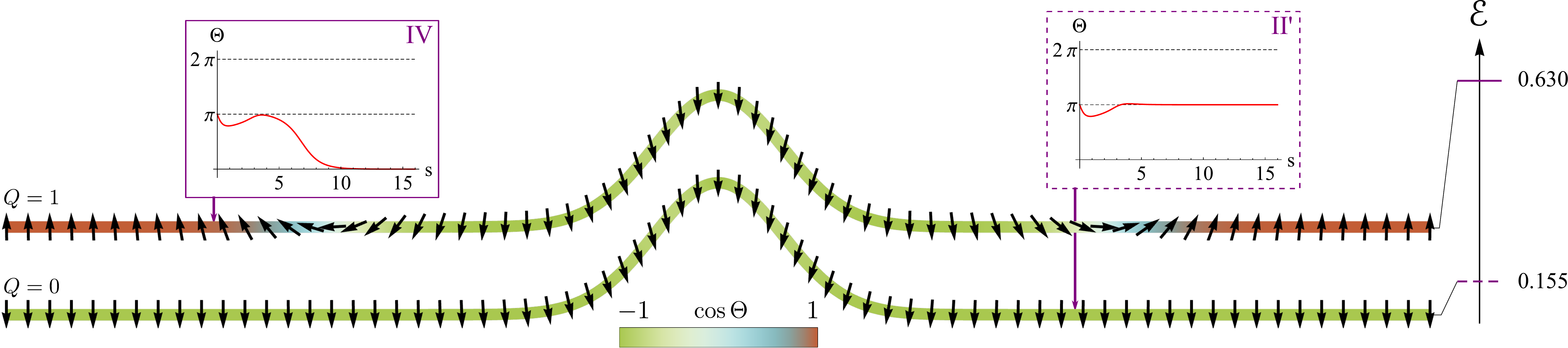}
	\caption{Spectrum of equilibrium magnetization states, which correspond to point ``$d$'' in Fig.~\ref{fig:diagram} --- $\mathcal{A}=2$, $r_0=1$, $d=1.2$. Skyrmion IV and non-topological solution II$'$correspond to lines IV and II$'$ in Fig.~\ref{fig:energies}, respectively.}\label{fig:spectrum-d}
\end{figure*}

\begin{figure*}[h]
	\includegraphics[width=\textwidth]{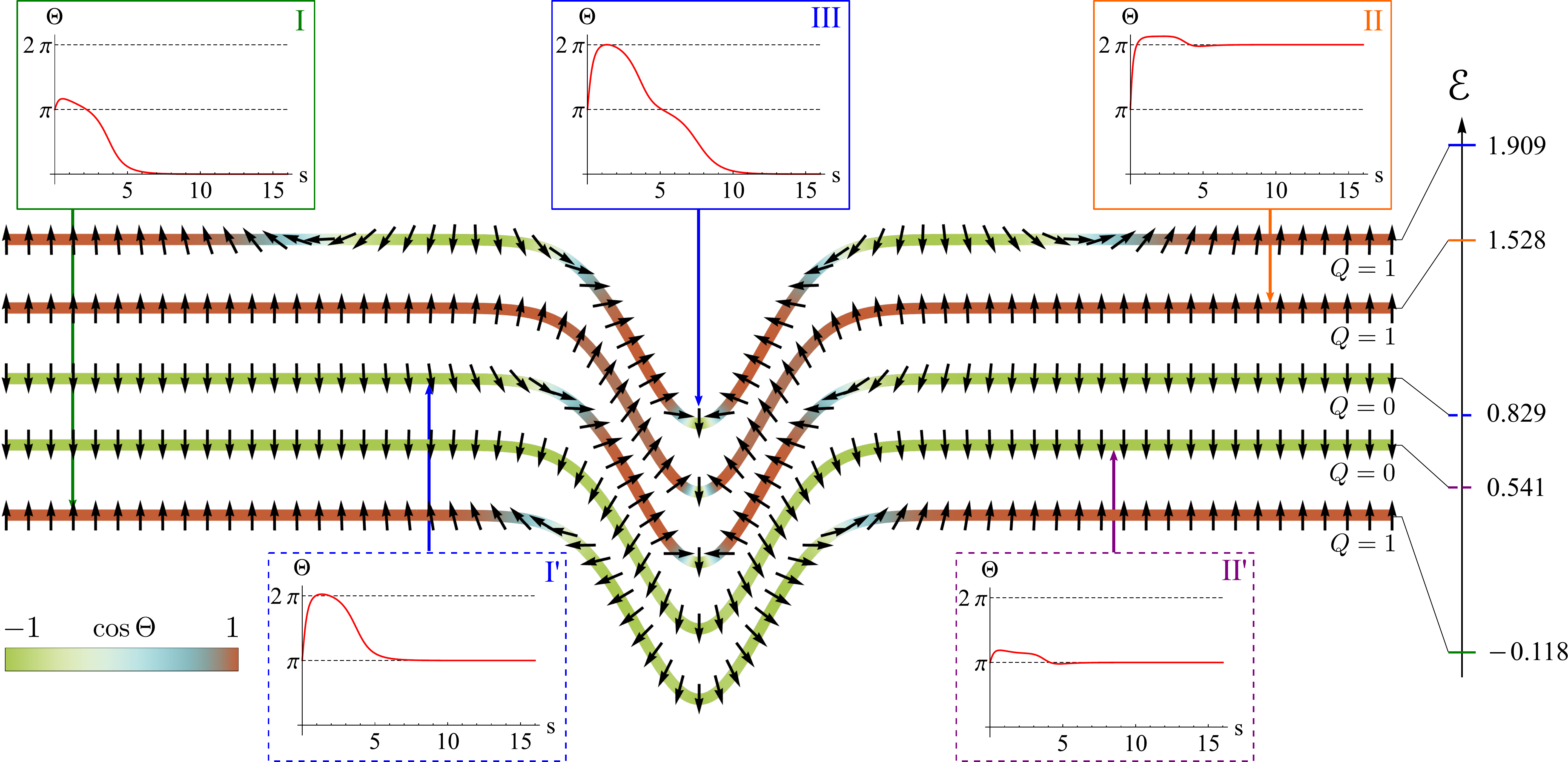}
	\caption{Spectrum of equilibrium magnetization states, which correspond to point ``$e$'' in Fig.~\ref{fig:diagram} --- $\mathcal{A}=-3$, $r_0=1$, $d=1.1$. }\label{fig:spectrum-e}
\end{figure*}

\begin{figure*}[h]
	\includegraphics[width=\textwidth]{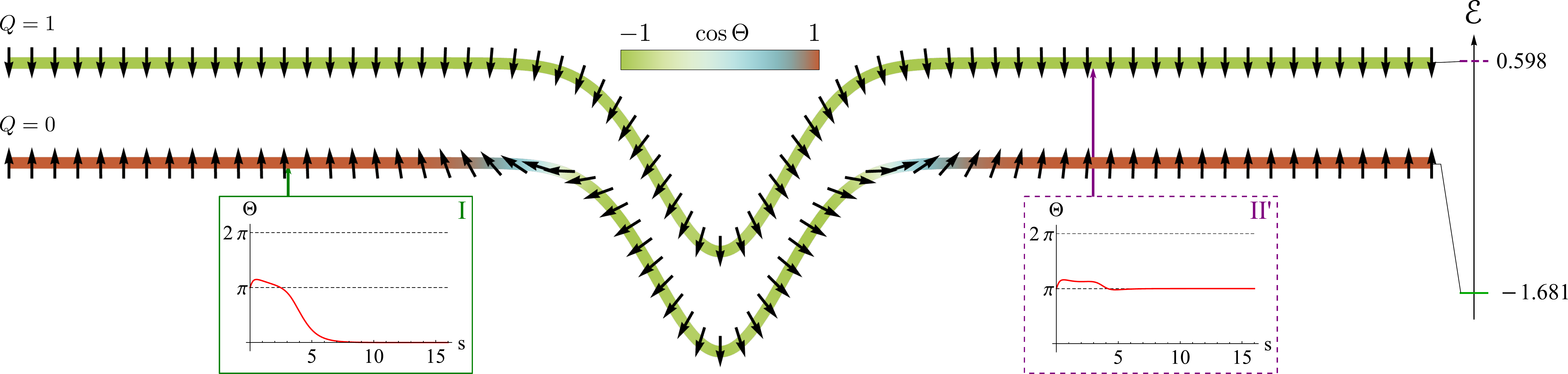}
	\caption{Spectrum of equilibrium magnetization states, which correspond to point ``$f$'' in Fig.~\ref{fig:diagram} --- $\mathcal{A}=-3$, $r_0=1$, $d=1.8$. }\label{fig:spectrum-f}
\end{figure*}
%

%

%

\end{document}